\input{aipcheck}

\documentclass[
    ,final            
  ]
  {aipproc}

\layoutstyle{6x9}

\begin{document}

\title{Typical AGN at intermediate redshifts}


\author{Almudena Alonso-Herrero}{
  address={DAMIR, Instituto de
  Estructura de la Materia, CSIC, 28006 Madrid, Spain}
}



\begin{abstract}
We study the host galaxies and black holes of
typical X-ray selected AGN at intermediate redshifts ($z\sim
0.5-1.4$). The AGN are  
selected such that their spectral energy distributions
are dominated by stellar emission, i.e., they show a prominent $1.6\,\mu$m
bump thus minimizing the AGN emission contamination. This AGN
population comprises approximately 50\% of the X-ray selected AGN at
these redshifts. AGN 
reside in the most massive galaxies at the redshifts probed
here, with characteristic stellar masses 
that are intermediate between those
of local type 2 AGN and high redshift ($z \sim 2$) AGN. 
The inferred black hole masses of typical AGN  are similar 
to those of optically identified quasars at similar redshifts. 
Since the AGN in our sample are much less luminous than quasars, typical AGN have 
low Eddington ratios. 
This suggests that, at least at intermediate redshifts, 
the cosmic AGN 'downsizing' is due to both a 
decrease in the characteristic stellar mass of the host galaxies,  and less
efficient accretion.
Finally there is no strong evidence in AGN host galaxies 
for either highly suppressed star
formation, expected if AGN played a role in quenching star formation, or
elevated star formation when compared to mass selected 
galaxies of similar 
stellar masses and redshifts. 
\end{abstract}

\maketitle


\section{Introduction}

One of the challenges faced by 
galaxy formation models is to explain the population of 
today's red massive quiescent elliptical galaxies. AGN feedback 
has been proposed as an efficient process for suppressing any
further star formation in the late stages of galaxy evolution, while 
still allowing for continuing black hole (BH) growth.
The location of
intermediate-$z$ X-ray selected AGN (\cite{nandra}) in the transition between 
the 'red sequence' and the top of the 'blue cloud' 
may indicate that AGN play a role in causing or maintaining the quenching of
star formation.  However, 
in the local universe AGN with strongly 
accreting BH tend to be hosted in massive galaxies with blue 
(i.e., star-forming) disks and young bulges (\cite{kauff03,kauff07})
implying a close link between the growth of BH and bulges. 

About half of AGN with  $L_{\rm X} > 10^{41}\,{\rm erg \, s}^{-1}$  
at intermediate-$z$ do not show broad lines or high 
excitation lines characteristic of AGN in their 
optical spectra (\cite{szokoly}). 
Since the AGN emission does not dominate 
their rest-frame UV to near-infrared (NIR) emission (\cite{rigby}), 
they are the ideal targets to study
their host galaxies and investigate the
role of AGN in galaxy evolution.  In this paper we study the 
host galaxies of X-ray selected 
AGN with stellar dominated spectral energy distributions (SEDs) at
intermediate redshifts ($0.5<z<1.4$) in the  
{\it Chandra} Deep Field South (CDF-S). The AGN host galaxy properties are 
then compared with a sample of stellar-mass selected galaxies 
(\cite{pg08}). 
Full details of the study are discussed by \cite{aah08}. 
We assumed: $H_0=70\,{\rm km \, s}^{-1} \, {\rm
  Mpc}^{-1}$, $\Omega_{\rm M=}0.3$ and  
$\Omega_\Lambda=0.7$.

\section{Sample, observations and modelling of SEDs}
We started with all the CDF-S X-ray sources  with redshifts in the
range of $0.5<z<1.4$. Then we restricted ourselves to the 
GOODS CDF-S field. We cross-correlated the positions of the X-ray sources
with the IRAC (simultaneous detections at $3.6$ and $4.5\,\mu$m)-selected 
galaxies of \cite{pg08}. We constructed 
the SEDs using the photometric catalogs of \cite{pg08} which include 
the two other IRAC bands, UV, optical, NIR,  and {\it
  Spitzer}/MIPS $24\,\mu$m data. 
For this study 
we selected AGN with stellar-dominated UV through NIR SEDs, 
and in particular with a strong $1.6\,\mu$m bump resulting in a sample
of 58 AGN. This selection minimized the
AGN contamination which is essential for studying the properties of their host galaxies. Our selection  thus excluded 
X-ray sources with AGN-dominated SEDs such as IR
power-law galaxies (\cite{aah06,donley}) or  
IRAC color-color selected AGN (\cite{lacy,stern}).
More importantly 
AGN with stellar dominated SEDs comprise approximately 50\% of the
population of X-ray selected AGN. 
52 AGN in the sample are optically-dull, that is, they do
not broad or high excitation emission line characteristic of AGN. The
remaining 6 AGN are optically-active, that is, they have high
excitation emission line. 
We fitted their rest-frame UV through MIR SEDs using stellar and dust models to
derive the stellar masses ($\mathcal{M}_*$) 
as well as the total (UV+IR) star formation rates (SFRs).  
The stellar masses were calculated for a Salpeter IMF
between 0.1 and 100\,M$_\odot$. The SED modelling also allowed us to
obtain photo-$z$  for the 6 AGN in our sample without
spectroscopic redshifts.  \cite{pg08,aah08}
 give full details of the modelling.


\section{Properties of typical AGN at intermediate-z}

\subsection{Stellar masses of the host galaxies}
Fig.~1 shows the redshift evolution of the stellar mass of 
AGN compared with that of the 
IRAC-selected sample of galaxies of \cite{pg08}. 
At the redshifts probed here the IRAC-selected comparison sample is essentially a
stellar mass selected sample. Clearly
 (X-ray identified) AGN reside in galaxies with 
a range of about an order of magnitude in mass, including some among the 
most massive at these intermediate redshifts. 
The characteristic stellar masses are $7.8 \times 10^{10}\,{\rm M}_\odot$  
at $0.5<z<0.8$ (median
$z=0.67$) and $1.2 \times 10^{11}\,{\rm M}_\odot$ 
at $0.8<z<1.4$ (median $z=1.07$). 
These masses are intermediate between those
of local type 2 AGN (\cite{kauff03}) and $z \sim 2$ AGN (\cite{kriek}). 
The evolution of  the quenching
mass (mass above which, star formation  should be mostly suppressed)
inferred by \cite{bundy} is also shown in Fig.~1. The fraction of
AGN above the line  is small perhaps indicating that star formation has not
been fully suppressed yet in these galaxies. It is important to stress that
the  AGN studied here comprise $\sim 50\%$ of the X-ray selected 
AGN population at $0.5 < z < 1.4$. Ideally we would like to estimate 
the stellar masses for all  optically-active AGN at intermediate redshifts,
but this   becomes increasingly 
more  uncertain, 
as for more luminous X-ray sources  
the AGN emission in the optical-NIR becomes more dominant
(\cite{donley}). The masses of 
the six optically-active AGN in our sample do not appear to be fundamentally different from
optically-dull AGN, although we note that the number statistics is
small.

\begin{figure}
  \includegraphics[height=.2\textheight]{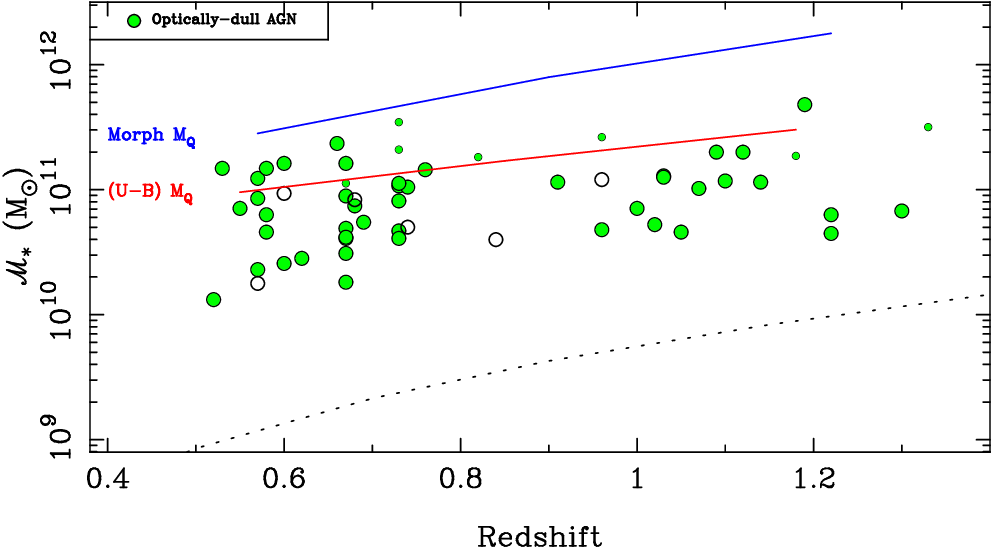}
  \includegraphics[height=.2\textheight]{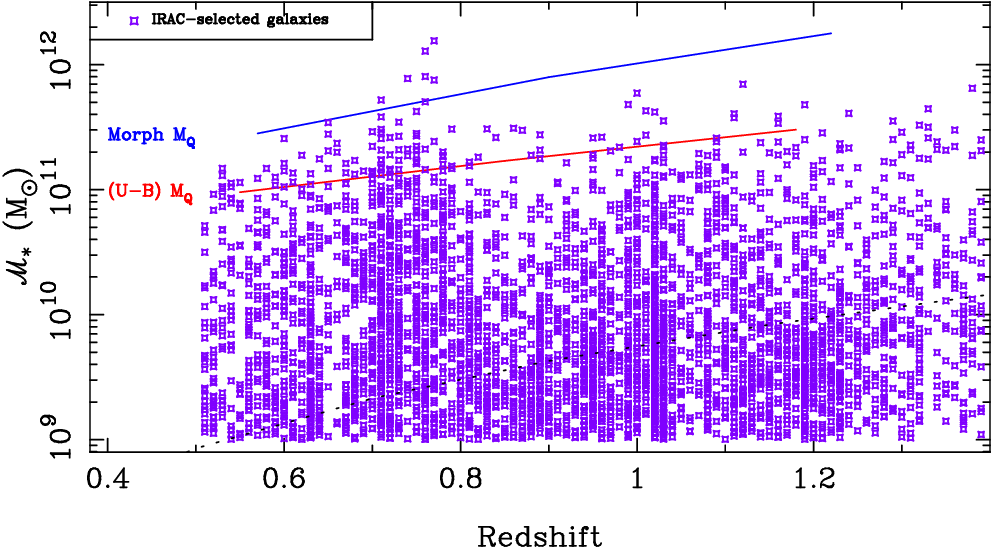}
  \caption{{\it Left panel:} 
Redshift evolution of the stellar mass of CDF-S AGN 
(all circles). 
The open circles are the stellar masses for  six optically-active AGN. 
The dotted line indicates the completeness limit of the sample of 
\cite{pg08} for a maximally-old passively evolving
galaxy. The two  
solid lines are two empirical determinations (color and morphology) of the 
redshift evolution of the quenching mass (converted to a
Salpeter IMF) 
of \cite{bundy}. The solid 
lines reflect two different criteria used by \cite{bundy} to 
evaluate star formation: color and morphological type.  
{\it Right panel:}
  Redshift evolution of the stellar mass of IRAC-selected galaxies
  (\cite{pg08}) in the CDF-S. Only galaxies with $\mathcal{M}_*>10^{9}\,{\rm
    M}_\odot$ are plotted in this comparison.
}
\end{figure}

\subsection{Star formation activity}

 In this section we quantify the star
formation activity of AGN with stellar-dominated SEDs in relation to IRAC-selected
galaxies of similar stellar masses and at similar redshifts. 
We use the specific SFR ($={\rm SFR}/\mathcal{M}_*$) 
as an indicator of the star formation activity rather
than the absolute SFRs, as the specific 
SFR measures the rate at which 
new stars add to the assembled mass of the galaxy. 
Fig.~2 (left panel) compares the median specific SFRs for
intermediate-$z$ AGN   with those of the
IRAC-selected sample of \cite{pg08}.
The AGN specific SFRs do not appear to be fundamentally
different from those of IRAC-selected galaxies. 
In contrast,  \cite{kriek} found evidence for a relation between the suppression of star
formation and the AGN phase for $K$-band selected galaxies. 
In the local universe AGN tend to be hosted in
massive galaxies with younger stellar ages than non-AGN  of similar
morphological types (early-type) and stellar masses (\cite{kauff03,kauff07}). This was 
interpreted  as 
evidence that enhanced star formation is a requisite for
feeding the AGN. At $z\sim 1$ galaxies (presumably both AGN and non-AGN) with 
$\mathcal{M}_*  \sim 10^{10}- 5 \times 10^{11}\,{\rm M}_\odot$ are still being
assembled (see \cite{pg08}), so
perhaps it 
is not surprising that the SFR of AGN and non-AGN are not 
significantly different.

\begin{figure}
  \includegraphics[height=.3\textheight]{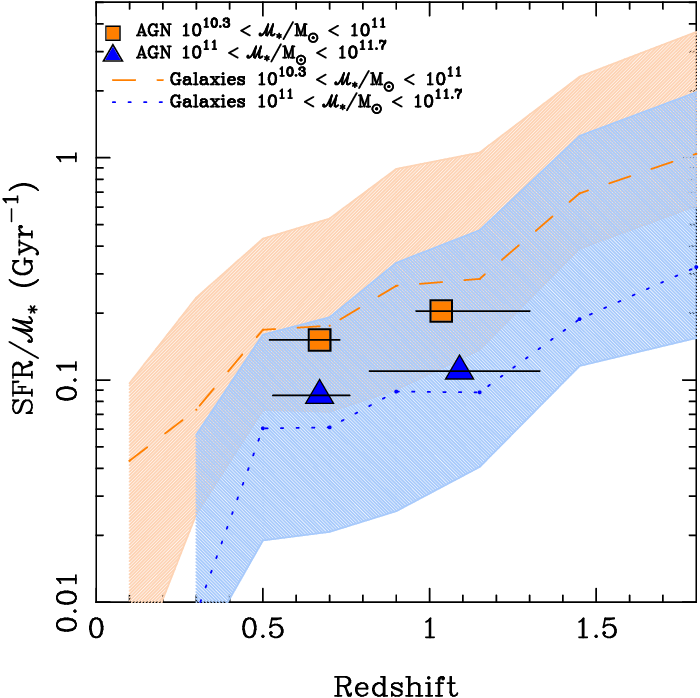}

  \includegraphics[height=.3\textheight]{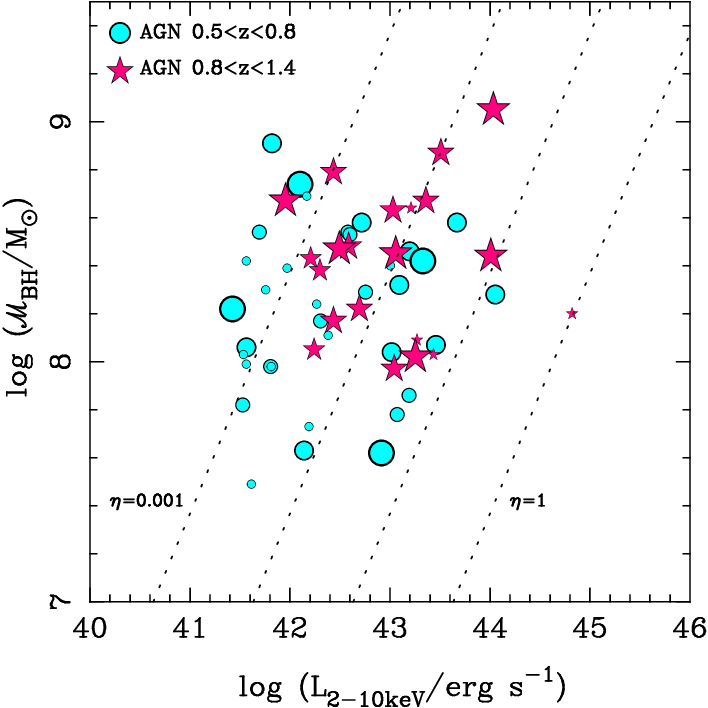}
  \caption{{\it Left panel:} Redshift evolution of the specific SFR for IRAC-selected 
galaxies for the cosmological fields studied by \cite{pg08}. We only
show here the 
mass intervals of interest for our AGN sample. 
The dashed and dotted lines are the median specific SFR for 
 the ranges $\mathcal{M}_* = 2\times 10^{10} -10^{11}{\rm M}_\odot$ 
and $\mathcal{M}_* 
= 1-5 \times 10^{11}{\rm M}_\odot$, respectively. 
We also show the quartiles of the distribution of specific SFRs of 
IRAC-selected
  galaxies for each of the two mass ranges as the 
shaded regions. The median specific SFRs for AGN
  are the filled squares and triangles plotted at the median redshift. The
  horizontal bars represent the ranges of redshifts.  
{\it Right panel:} BH masses  
versus absorption-corrected rest-frame 
hard X-ray luminosities. The circles are AGN at $0.5 < z < 0.8$ and the
star symbols are AGN at $0.8 < z < 1.4$. 
The sizes of the symbols are proportional to their 
rest-frame $24\,\mu$m luminosities. 
The dotted lines indicate from right to left 
Eddington ratios of  $\eta =1, 0.1, 0.01$ and 0.001.}
\end{figure}

\subsection{Black holes masses and Eddington ratios}

Using the local relation between 
$\mathcal{M}_{\rm bulge}$ 
and $\mathcal{M}_{\rm BH}$ of \cite{marconi} the inferred 
BH masses have a median value of $\mathcal{M}_{\rm BH} \sim 2 \times
10^{8}\,{\rm M}_\odot$.  These BH masses are similar to
those of broad-line AGN although with lower Eddington ratios ($\eta \sim
0.01-0.001$, see Fig.~2, right panel) than luminous quasars. 
One possibility to explain the low Eddington ratios is that we were 
overestimating significantly 
the BH masses.  This does not seem to be the case as the host galaxies
appear to be 
bulge-dominated (\cite{ballo}), 
so the assumption $\mathcal{M}_{\rm bulge}\simeq
\mathcal{M}_*$ is probably correct.
The low Eddington ratios together with the characteristic stellar masses of the host
galaxies of AGN suggest that, at least at intermediate-$z$, 
the cosmic AGN 'downsizing' is probably due not only to a 
decrease in the characteristic stellar mass of the host galaxy
(\cite{heckman}), 
but also to less efficient accretion (see also \cite{babic}).



\end{document}